# Plasticity of an extra-strong nanocrystalline stainless steel controlled by the "dislocation-segregation" interaction

N.A. Enikeev[1,2,*], I.V. Lomakin[2,3], M.M. Abramova[1], A.M. Mavlyutov[2], A.A. Lukyanchuk[4], A.S. Shutov[4], X. Sauvage[5]

[1] *Institute of Physics of Advanced Materials, Ufa State Aviation Technical University, 450008 Ufa, Russia*
[2] *Laboratory of Mechanics of Advanced Bulk Nanomaterials for Innovative Engineering Applications, Saint Petersburg State University, 198504 Peterhof, Saint Petersburg, Russia*
[3] *Department of Applied Physics, Aalto University, 00076 Aalto, Espoo, Finland*
[4] *Institute for Theoretical and Experimental Physics, NRC Kurchatov Institute, 117218 Moscow, Russia*
[5] *Normandie Université, UNIROUEN, INSA Rouen, CNRS, Groupe de Physique des Matériaux, 76000 Rouen, France*

[*]corresponding author: nariman.enikeev@ugatu.su

**Abstract:** We study three structurally different states of nanocrystalline 316 steel and show that the state, where boundaries containing excess concentration of alloying elements are combined with mobile dislocations in grain interiors, allows maintaining extraordinarily high strength and remarkably enhanced plasticity. Underlying mechanisms featuring interaction between the segregations and mobile dislocations are discussed.

**Introduction**

Increasing yield strength of austenite stainless steels is a topical issue to boost up their potential application, in particular, for light-weight solutions in medicine [1]. Formation of nanocrystalline (NC) states in these steels allows skyrocketing their strength accompanied with significant decrease of elongation to failure [2-4]. Interestingly, annealing at certain temperatures leads to additional strengthening and further reduction in plasticity [2,3] of the NC 316 steel even though the dislocation density decreases, and no precipitation occurs. Such a hardening on annealing effect observed also for the other nanostructured metals was attributed either to dislocation starvation mechanism [3,5-8] or to formation of strain-induced grain boundary (GB) segregations [2,4]. Huang et al [5] and the following studies carried out on a number of Al alloys demonstrated a striking effect of additional straining on the plastic behavior of a nanostructured material subjected to the anneal hardening treatment (see the recent results reported in [6] and citations ibid). This effect was explained by introducing strain-induced dislocations which promote deformation.

We propose an alternative vision of tailoring plasticity in the extra-strong NC steel by controlling interaction of dislocations with GB segregations.

**Materials and Methods**

Conventional coarse-grained stainless 316 steel [4] was nanostructured by high pressure torsion (HPT) by straining to a steady state. Three NC states were prepared:

(i) Nanostructured at room temperature (**as-deformed**)

(ii) The previous state annealed at 550°C for 1h (**annealed**)

(iii) The previous state followed by small extra-deformation by HPT (1/4 rotation) at room temperature (**extra-strained**)

To reveal fine microstructural features, Atom Probe Tomography (APT) analyses were carried out with the APPLE-3D device [9]; TEM and XRD studies were performed using JEOL JEM 2100 microscope and Bruker D8 diffractometer, respectively. Tensile tests were carried out at room temperature with Shimadzu AGX-50 Plus machine adjusted for testing mini-specimens. The details of the experimental procedures are specified in supplementary materials.

**Results and discussion**

Deformation curves are presented in Fig. 1 with values for yield stress ($\sigma_{02}$), ultimate tensile strength ($\sigma_{UTS}$) and total elongation to failure ($\delta$) listed in Table 1. These data indicate striking differences in mechanical performance of 316 steel among studied NC states. The strength of the as-deformed steel exhibits many-fold growth, while the elongation to failure reduces to several percent. Annealing leads to further considerable increase in strength together with abrupt decrease in $\delta$. Extra-straining remarkably enhances plasticity while the strength is maintained closer to the level of the annealed steel.

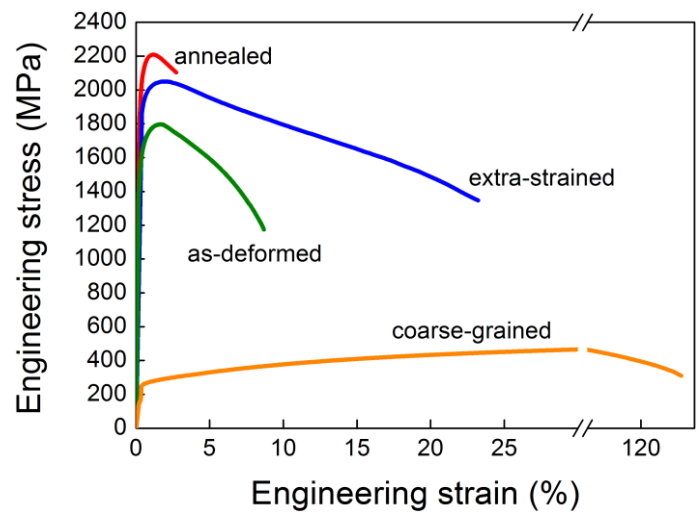

Fig. 1. Tensile engineering curves of 316 stainless steel in different NC states

Table 1. Mechanical properties of 316 steel in different structural states

| state | $\sigma_{02}$ (MPa) | $\sigma_{UTS}$ (MPa) | $\delta$ (%) |
| --- | --- | --- | --- |
| Coarse-grained | 258±6 | 528±5 | 120±6 |
| As-deformed | 1705±24 | 1810±26 | 7±1 |
| Annealed | 2100±41 | 2150±32 | 4±1 |
| Extra-strained | 1888±31 | 2018±43 | 26±3 |

Fig. 2 presents TEM observations. As it has been also shown earlier [4,10] HPT leads to complete nanostructuring of 316 steel to an average effective grain size of 40-50 nm (Fig. 2a). Stainless steels could undergo HPT-induced phase transformations at room temperature [11]. However, as-deformed specimens exhibited minor content of martensite, below XRD detection limit and only rare weak spots of $\alpha$-phase were observed at some TEM diffraction patterns. Accordingly, phase transformation effects were neglected in further considerations.

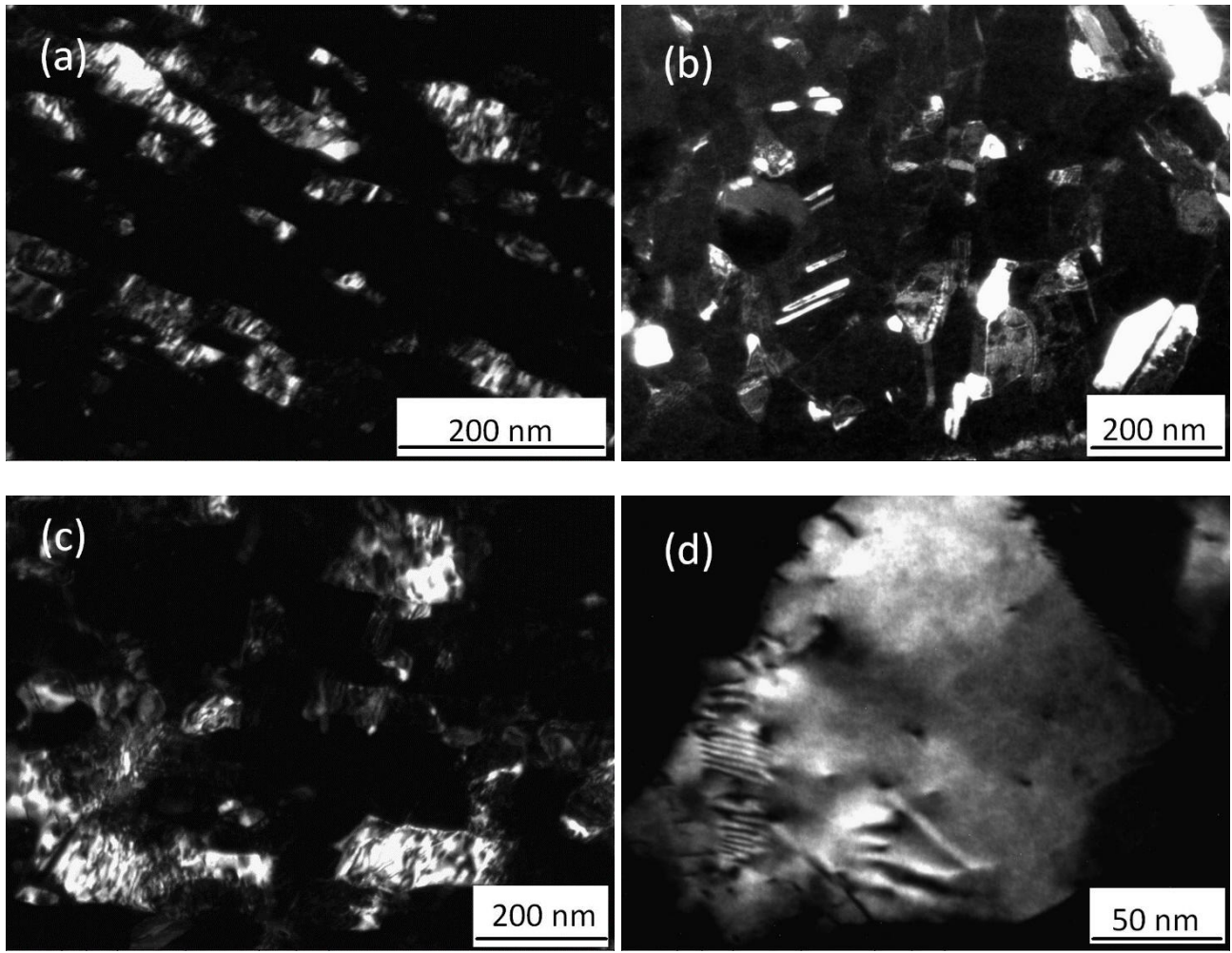

Fig. 2. Dark-filed TEM images of NC 316 steel: the as-deformed (a), annealed (b) and extra-strained states (c); a magnified area of the extra-strained state (d).

Fig. 2b testifies that annealing of the as-deformed state does not provide noticeable grain growth. It shows that heat treatment recovers the distortions inside grains meaning that the intragranular dislocation density notably decreases. Extra-straining forms new distortions inside grains (Fig. 2c,d) indicating that dislocations were activated and stored again. The magnified TEM image presented in Fig. 2d indicates preferable location of the dislocations in the GB area – for the illustrative purpose this figure displays a particular grain having larger size than the average one to resolve the difference in intragranular dislocation configurations among the annealed and extra-strained states. TEM results are well consistent with the XRD measurements taken to quantify the difference in dislocation densities among the three states. The as-deformed, extra-strained and annealed states have similar coherent domain sizes of ~20 nm and the highest ($2.5\times10^{15}$ m$^{-2}$), intermediate ($1.8\times10^{15}$ m$^{-2}$) and lowest ($0.77\times10^{15}$ m$^{-2}$) dislocation density, respectively.

As previously reported [3,12] HPT at room temperature does not entail re-distribution of alloying elements in 316 steel. Fig. 3a shows that annealing leads to a gradient increase of Si, Mo and Cr concentration in the GB area accompanied with corresponding depletion in Fe, which is consistent with earlier reports [3,10].

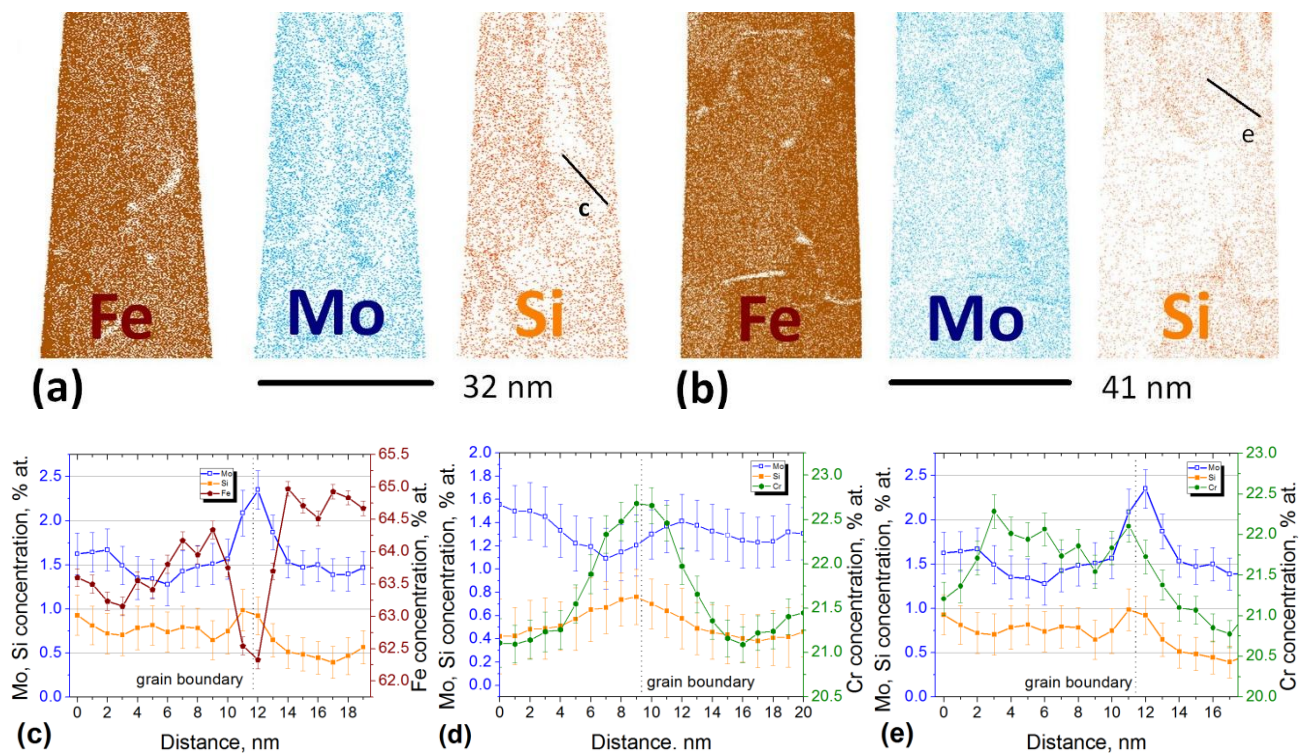

Fig. 3. The reconstruction of atomic structure by APT analyses for (a) annealed and (b) extra-strained states of NC 316 steel, where Fe is denoted by brown, Mo by blue and Si by orange colors. Concentration profiles across segregated GBs revealing heterogeneity in re-distribution of solutes

(d,e) accompanied with depletion boundaries in Fe (c). The profiles (c,e) were collected along the lines denoted in figures (a,b)

Note, that these segregations are heterogeneous, moreover, some boundaries are depleted with the same solutes. This can be explained by the microstructure inherited after large deformation having highly non-equilibrium and heterogeneous nature [13] and by competition of different elements for the space in interfaces. Fig. 3b shows that extra-straining does not entail any qualitative changes in the local atomic structure of the annealed NC steel: boundaries decorated with the same solutes are clearly visible. No significant difference in the excess concentration of alloying elements in GBs have been found among these two states. Concentration profiles (Fig. 3c-e) confirm that the appearance of GB segregations is beyond the measurement error.

Our findings suggest that interaction of GBs, segregations and dislocations can be an important micro-mechanism governing plasticity in NC alloys. To date, the effect of *extra-straining* in annealed fine-grained materials was explained by introducing extra-density of dislocations improving plasticity and reducing the yield stress [5,6]. However, the presented results allow revisiting the phenomenon of "hardening by annealing and softening by deformation" [5]. It is natural to propose an alternative mechanism (schematically presented in Fig. 4) for the drastic improvement in plasticity in the high strength state of NC 316 steel with possible expanding this vision to the other fine-grained alloys.

It is well known, that in nanocrystalline materials the dominant deformation mode can change from dislocation slip to emission of partials or/and grain rotation, which depends on stacking fault energy, strain rate, temperature and so on. In our study we assume that the main deformation mode in our case is controlled by dislocation slip which is confirmed by the texture measurements of the deformed 316 steel. Maxima of pole density on the (111) pole figure correspond well to the ideal orientations of simple shear (see Fig. S1b in the supplementary materials).

Dislocations are intensively forming and interacting during HPT so that the resulting deformed state contains dislocations retaining in the grain interiors and grain boundaries. During tensile test of the deformed state dislocations are both nucleated/emitted from GBs (which requires less stress in the absence of GB segregations) and slipping in the grain interiors (Fig. 4a).

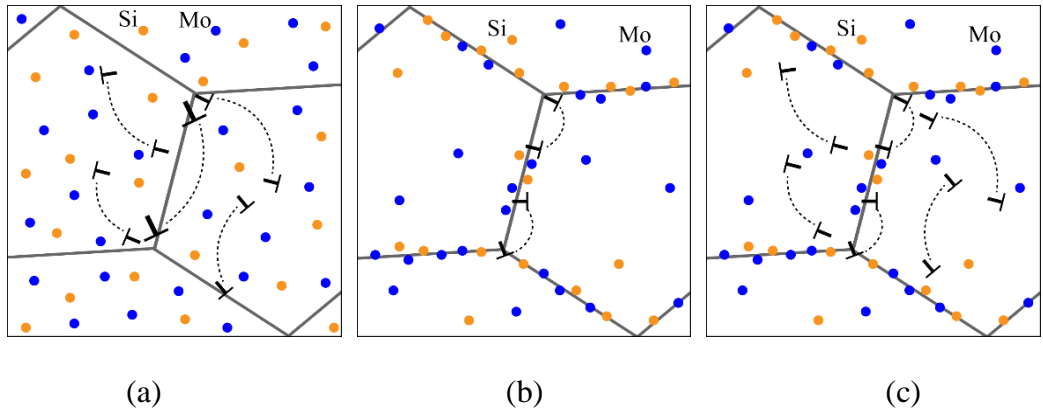

(a)          (b)          (c)

Fig. 4. A probe boundary of the NC 316 steel with different configurations of interacting defects. Blue and orange circles represent Mo and Si atoms, respectively, black lines – dislocations. The figure schematically illustrates the as deformed (a), annealed (b) and extra-strained (c) cases

Assuming high level of retained dislocation density in the annealed NC steel, the reduced amount of visible dislocations does not mean their total annihilation, but rather their recombination to GB area where they are pinned by solute clusters (Fig. 4b). In order to unpin the dislocations and to start plastic flow, an extra-stress is required. This stress can be defined by the reduced length of the nucleating [14] or emitting dislocation segments [15] or by mismatch in a lattice parameter between the segregated inclusion and a matrix [16]. Since pinned dislocations are located in GBs, deformation of the annealed state leads to fast accumulation of critical stresses in the localized areas. This results in the early intergranular fracture, providing high yield stress but significant reduction in elongation. Introducing additional density of mobile dislocations into the annealed state (Fig. 4c) provides more uniform distribution of internal stresses as well as enhances propagation of deformation by unpinned dislocations via intragranular slip under applied external stress, while remaining GB segregations maintain high barriers for the nucleation and emission of dislocations keeping high flow stress.

**Conclusion**

We deduce two major conclusions: (i) the elongation to failure of the extra-strong NC 316 steel can be significantly improved based on the new vision of the deformation micro-mechanisms in fine-grained alloys; (ii) we propose an alternative mechanism controlling the plasticity based on the interaction of heterogeneous GB clusters/segregations with mobile dislocations. This consideration expands current understanding of the fundamental features controlling plastic behavior of NC alloys.

**Acknowledgements**

The research was partially supported by the Russian Science Foundation (project #20-63-47027). The APT analyses were carried out using the equipment of the KAMIKS Center for Collective Use (CCU), Institute of Theoretical and Experimental Physics, National Research Center «Kurchatov Institute», Moscow. XRD studies were conducted in the CCU "Spectrum" (Institute of Molecule and Crystal Physics) and the regional CCU "Agidel" of Ufa Federal Research Center RAS.